\documentclass[aip,jcp,amsmath,amssymb,preprint]{revtex4-1}

\usepackage{graphicx}
\usepackage{dcolumn}
\usepackage{bm}

\draft 

\begin{document}

\preprint{J. Chem. Phys.}

\title{On Calculation of Thermal Conductivity from Einstein Relation in Equilibrium MD} 

\author{A. Kinaci}
\email{akinaci@tamu.edu}
\affiliation{%
Material Science and Engineering, Texas A\&M University, College Station, TX 77845-3122, USA}%
\author{J. B. Haskins}
\email{haskinjb@tamu.edu}
\author{T. {\c C}a{\u g}{\i}n}%
\email{tcagin@tamu.edu}
\affiliation{%
Artie McFerrin Department of Chemical Engineering \& Material Science and Engineering, Texas A\&M University, College Station, TX 77845-3122, USA}%



\date{\today}

\begin{abstract}

In equilibrium molecular dynamics, Einstein relation can be used to calculate the thermal conductivity. This method is equivalent to Green-Kubo relation and it does not require a derivation of an analytical form for the heat current. However, it is not commonly used as Green-Kubo relationship. Its wide use is hindered by the lack of a proper definition for integrated heat current (energy moment) under periodic boundary conditions. In this paper, we developed an appropriate definition for integrated heat current to calculate thermal conductivity of solids under periodic conditions. We applied this method to solid argon and silicon based systems; compared and contrasted with the Green-Kubo approach.

\end{abstract}

\pacs{}

\maketitle 

\section{Introduction}
Accurate determination of thermal conductivity of materials is of fundamental importance especially for technological applications where the control of thermal transport is critical. For instance, microelectronics, where the device failures are mostly due to overheating, require a strict thermal management. The development of efficient thermoelectrics require materials and materials systems with low thermal conductivity. A similar requirement arise in the development of thermal barrier coatings. Experimental evaluation of thermal conductivity is not as straightforward as electrical conductivity. Measurements generally suffer from heat losses, non-uniform heating and errors introduced due to approximations that account for sample size and structure~\cite{Rowe_2006}. As the dimensions of the samples get smaller, these effects become more pronounced, increase the inaccuracy of the measurements.
Due to their wide use in electronics, thermal transport in silicon and silicon based alloys has been studied extensively. Within the last decade, silicon based materials have been considered for thermoelectric and thermal barrier applications where a small thermal conductivity is desired~\cite{Snyder_2008}. The studies in this area generally focus on the influence of defects, impurities and nano-structure  (superlattices~\cite{Dames_2004}, nanowires~\cite{Hochbaum_2008}, quantum dots~\cite{Balandin_1998}, etc.) on thermal transport.

As an alternative to experiments, computational approaches offer controllable virtual measurements to determine thermal conductivity, with the added potential of probing the processes and mechanisms that influence the thermal transport. The two most frequently used computational methods that include atomistic level details in studying thermal transport are based on Boltzmann transport equation (BTE) and molecular dynamics (MD) simulations. The first one, BTE, is a semiclassical approach. The atomistic level granularity in BTE, is provided by the determination of phonon frequencies and group velocities from the lattice dynamics of an N-body system. However, the detailed physics of the heat transfer is incorporated into the formulation through phonon relaxation times~\cite{Chen_2001_1,Chen_2001_2,Majumdar_1993}. This requires either experimental results or a detailed analysis of the fundamental scattering processes to provide reliable theoretical estimates of relaxation times. When such data is not available due to the complexity of the system, the relaxation times can be probed with MD. Details of such calculations are described in earlier studies where MD was used to obtain the phonon relaxation times in fcc crystals~\cite{Turney_2009} and silicon~\cite{Zhao_2008,Henry_2008} using the decay of phonon mode amplitudes or energies.

In contrast, the full atomistic detail of the physics of heat transport is inherently present in MD through the definition of interactions between the atoms that make up the system. The evolution of the state of the system is given by the equations of motion based on interaction  potentials that are usually anharmonic.
Hence, MD is a powerful tool, as it includes normal, Umklapp and higher order phonon-phonon scattering events through the dynamics of the thermodynamic state of the system. Of course due to the classical nature of MD, vibrational modes are equally excited (i.e. have the same energy $k_{B}T$),~\cite{Volz_2007,Cahill_2003} and the phonon scattering processes do not include electronic contributions. However, some approaches~\cite{Che_2000_1,Egelstaff_1962} have also implemented quantum corrections. There are three MD based approaches for calculating thermal conductivity: the steady state non-equilibrium (direct) method, synthetic hamiltonian non-equilibrium method and the equilibrium (Green-Kubo) method. The direct method mimics experiment by imposing a temperature gradient on the system and determines the thermal conductivity from the phenomenological Fourier law~\cite{Lukes_2000,Volz_1996,Chen_2004,Maiti_1997,Padgett_2006,Cem-NL}. Temperature gradients can be created by scaling the particle velocities, changing the ensemble averages or switching particle velocities between the hot and cold ends. There can be some complications with this method due to the nonlinear response of a system associated with very large temperature gradients. The second technique, proposed by Evans~\cite{Evans_1982} applies a homogenous synthetic field to create a heat flux, as opposed to applying a temperature gradient. Finally, Green and Kubo~\cite{Green_1954,Kubo_1957} showed that through the fluctuation-dissipation theorem, transport properties i.e. the diffusion coefficient ($\bm{D}$), shear viscosity ($\eta_{\mu\nu}$) and thermal conductivity ($\bm{\kappa}$) can be calculated by integration of autocorrelation functions of particle velocity ($\bm{v}^i$), pressure tensor ($P_{\mu\nu}$) and heat current ($\bm{J}$).


\begin{equation}
\bm{D}=\frac{1}{N}\int_0^\infty \bigg\langle \sum_{i=1}^N{\bm{v}^{i}(t)\otimes\bm{v}^{i}(0)} \bigg\rangle~dt
\label{eq:v_correlation}
\end{equation}

\begin{equation}
\eta_{\mu\nu}=\frac{1}{V k_B T}\int_0^\infty \bigg\langle {P_{\mu\nu}(t)P_{\mu\nu}(0)} \bigg\rangle~dt~~~(\mu\neq\nu)
\label{eq:p_correlation}
\end{equation}

\begin{equation}
\bm{\kappa}=\frac{1}{V k_BT^2}\int_0^\infty \bigg\langle \bm{J}(t)\otimes\bm{J}(0) \bigg\rangle~dt
\label{eq:correlation}
\end{equation}

where $\mu$ and $\nu$ are the vector components, $V$ is the system volume, $k_{B}$ is the Boltzmann constant, $T$ is temperature and $N$ is the total number of particles. The same transport coefficients may also been obtained via 'Einstein' relation; which are related to doubly integrated  form of these expressions.

\begin{equation}
\bm{D}=\frac{1}{N}\lim_{t \rightarrow \infty}\frac{1}{2 t}\bigg\langle\sum_{i=1}^N{[\bm{r}^i(t)-\bm{r}^i(0)]\otimes[\bm{r}^i(t)-\bm{r}i(0)]}\bigg\rangle
\label{eq:d_einstein}
\end{equation}

\begin{equation}
\eta_{\mu\nu}=\frac{1}{V k_B T}\lim_{t \rightarrow \infty}\frac{1}{2 t}\bigg\langle\bigg(\sum_{i=1}^N{p^{i}_{\mu}(t)r^{i}_{\nu}(t)-p^{i}_{\mu}(0)r^{i}_{\nu}(0)}\bigg)^2\bigg\rangle~~~(\mu\neq\nu)
\label{eq:p_einstein}
\end{equation}

\begin{equation}
\bm{\kappa}=\frac{1}{V k_BT^2}\lim_{t \rightarrow \infty}\frac{1}{2 t}\bigg\langle[\bm{R}(t)-\bm{R}(0)]\otimes[\bm{R}(t)-\bm{R}(0)]\bigg\rangle
\label{eq:einstein}
\end{equation}

In the Eqs.~\ref{eq:d_einstein},~\ref{eq:p_einstein} and~\ref{eq:einstein}, $\bm{r}^i$, $p_{i\mu}$ and $\bm{R}$ are the position and momentum of $i^{th}$ particle and integrated heat current respectively. The integrated heat current will be referred to as the energy moment in this paper and is defined for any N-body system as:

\begin{equation}
\bm{R}=\sum_{i=1}^N{\epsilon_{i}\bm{r}_i} \label{eq:Req}
\end{equation}
where $r_i$ is the position and $\epsilon_{i}$ is the energy content of the $i^{th}$ particle.

Detailed derivations on the time correlations functions and associated Einstein relations for transport properties can be found in several sources~\cite{Zwanzig_1965,McQuarrie_2000,Allen_1987}. In the case of thermal conductivity, the heat current autocorrelation function (HCACF) approach has been employed for a number of materials such as bulk diamond~\cite{Che_2000_1}, bulk silicon~\cite{Schelling_2002}, metal organic frameworks~\cite{Huang_2007}, nanotubes~\cite{Che_2000_2}, nanoribbons~\cite{Evans_2010} and nanoparticles~\cite{Domingues_2005}.
However, unlike the wide use of the Einstein relation in determining mass transport/diffusion, the use of the Einstein relation for thermal conductivity is rather rare.  This situation has arisen from a lack of proper implementation in the calculation of integrated heat current/energy moment in simulations.

In the application of the Einstein relation based thermal conductivity calculation in MD, discrepancies arise regarding the calculation of the energy moment and energy current. The techniques used for calculating $\bm{R}$ give an expression which produces a bounded behavior and results in zero thermal conductivity in non-diffusive solid systems, while $\bm{J}$ yields a reliable result. Similar problems have been reported for the calculation of shear viscosity from the Einstein relation using MD with periodic boundary conditions~\cite{Allen_1993,Rapaport_2004}. Here, we propose an expression for $\bm{R}$ that gives a proper thermal conductivity and can be reduced numerically to give the exact value of $\bm{J}$. Further, for solid systems, the resulting energy moment expression can be generalized to be independent of the potential model. In this report, we formulate the new method and demonstrate its use through applications to a two-body potential (Lennard-Jones) and then an N-body  potential (Tersoff)~\cite{Tersoff_1989}.

The paper is organized as follows: In the next section we present the background and reformulated form of the energy moment based method of calculating thermal conductivity. In section III, we present the application of the method, to Argon, Silicon, and Porous Silicon and silicon Germanium nanodots in Silicon matrix, providing a detailed comparison of energy moment based method with HCACF based methods. Where we observe that HCACF of heterogeneous systems (Si-Ge) or systems with extended defects (Si with pores) are showing large fluctuations that one needs to filter out or large variance in resulting thermal conductivity values.  In contrast the Einstein relation for all systems has a much better behavior. In section IV, we summarize our results and present our conclusions.

\section{Background and Theory}

The fundamental variable for calculating thermal conductivity from Eq.~\ref{eq:einstein} is the energy moment, $\bm{R}$ is defined above in Eq.~\ref{eq:Req}. The energy content of particle $i$, $\epsilon_{i}$, is defined as:

\begin{equation}
\epsilon_i=\frac{1}{2}m_iv_i^2+ u_i \label{eq:particle_energy}
\end{equation}

Here, the first and the second terms are the total kinetic and potential energy of the particle $i$. The velocity and the mass of the same particle are represented by $v_i$ and $m_i$. The heat current, $\bm{J}$, is the time derivative of $\bm{R}$.

\begin{equation}
\bm{J}=\frac{d\bm{R}}{dt} \label{eq:hc}
\end{equation}

As the general definition of $\bm{R}$ and $\bm{J}$ are given, we will describe the method of determining thermal conductivity with two-body potential and generalize to n-body potential. For two-body interactions, the total potential energy content of a particle is defined as $u_i=\frac{1}{2}\sum_{j\neq i}{u_{ij}}$ where $u_{ij}$ is the potential between particle $i$ and $j$. Then, the corresponding heat current can be calculated through the following expression:

\begin{equation}
\bm{J}=\sum_{i=1}^N{\epsilon_{i}\bm{v}_{i}}+\frac{1}{2}\sum_{i,j~i\neq j}{\bm{r}_{ji}(\bm{f}_{ij}^{i}\cdot{\bm{v}_{i}})}
\label{eq:hcacf}
\end{equation}

where $r_{ji}$ is defined as the nearest image distance and $f_{ij}^i$ the interatomic force on $i$ due to $j$.


A straightforward implementation of Eq.~\ref{eq:einstein} through Eq.~\ref{eq:particle_energy} we call \emph{formulation 1}, by considering a bulk non-convective solid in a periodic box, as we will show produces no thermal conductivity. Moreover, if $r_{ji}$ in Eq.~\ref{eq:hcacf} is the nearest image distance, then Eq.~\ref{eq:Req} cannot be reduced to Eq.~\ref{eq:hcacf}. The issue of defining $\bm{R}$ in periodic systems has been pointed out by Donadio and Galli~\cite{Donadio_2010}, who attributed the inconsistency to the ill-definition of particle positions in a periodic system. However, we believe this is a consequence of the omission of what we call cross-boundary interactions, interactions or transfer of energy between the $N$ atoms in the simulation region and the image atoms across the periodic boundary. These interactions are vital to accurately defining the energy dispersion of the system and must be included into the definition of $\bm{R}$. With the inclusion of the cross-boundary interactions, energy moment can be separated into potential ($\bm{R}_p$) and kinetic ($\bm{R}_k$) contributions. The combination of $\bm{R}_p+\bm{R}_k$ will be called \emph{formulation 2} for the rest of the report.

The new form for the potential energy portion of $\bm{R}$, $\bm{R}_p$, is
\begin{equation}
\bm{R}_p=\sum_{i,j>i}\frac{u_{ij}}{2}(\bm{r}_{i}+\bm{r}_{j})\label{eq:Rp}
\end{equation}
where $u_{ij}$ is the pair potential, $i$ represents atoms in the box and $j$ represents the (real or image) neighbors of $i$. To avoid double counting, $j>i$ constraint is imposed. Though the form is very similar to Eq.~\ref{eq:Req}, if an atom in the simulation box interacts with an image atom, the actual atomic position of the image is used, instead of contributing the image interaction to its corresponding particle in the box. Although a simple equipartition of potential energy is the reasonable choice in the case of two-body interactions, n-body interactions may require other treatments and a rule of energy distribution may not preferable to another. In these circumstances, one needs to derive a separate expression for the heat current from $\bm{R}$ for each choice of potential splitting. However, this derivation is avoided when using $\bm{R}_p$ directly.

The kinetic portion of the energy moment requires a similar approach. In short, the transfer of kinetic energy from atoms in the box to image atoms must be included. This transfer is mediated by power term, $\bm{f}\cdot\bm{v}$. The kinetic portion, $\bm{R}_k$, can then be written as:
\begin{equation}
\bm{R}_k=\sum_{i,j>i}(\bm{r}_{i}\int_0^t\bm{f}_{ij}^i\cdot\bm{v}_{i}\ dt+\bm{r}_{j}\int_0^t\bm{f}_{ij}^j\cdot\bm{v}_{j}\ dt) \label{eq:rk1}
\end{equation}
where $\bm{f}_{ij}^i$ represents the force on atom $i$ due to $j$. This equation represents the transfer of kinetic energy from atom $i$ to all particles that interact with it. Again, if the neighbor is an image atom, the image position is used.

This discussion has been carried out for the two body interaction, but it can also be formatted for n-body potential. In fact, the kinetic part can be put in a form that does not depend explicitly on the potential form by simply rearranging the summation as:
\begin{equation}
\bm{R}_k=\sum_{l}\bm{r}_{l}\int_0^t\bm{f}_{l}\cdot\bm{v}_{l}\ dt \label{eq:rk}
\end{equation}
where $l$ is a summation over particles (real or image) that make up unique n-body interaction groups. Though similar to the previous expression, Eq.~\ref{eq:rk} is well defined for any interatomic potential as $\bm{f}_{l}$ is simply the force on particle $l$ due to all unique interaction groups that include $l$.

The computations can be restricted to the kinetic portions of the energy moment when dealing with the non-convective systems. In solids, there is very little convection, and to see diffusion in a simulation is highly improbable. This, along with the fact that the potential contribution to the atomic energy is bound, leads to the conclusion that $\bm{R}_p$ does not contribute to the thermal conductivity in perfect solid systems. Then, $\bm{R}_k$ must be main component driving the increase in the Einstein relation. On the other hand, for highly diffusive systems such as liquids, $\bm{R}_p$ is expected to contribute significantly.

\section{Results and Discussions}

\subsection{Model Systems}

In order to further illustrate the formulations given in the theory, we studied archetypical two-body (argon) and n-body (silicon) systems. Moreover, we discuss the behavior of Einstein relation when impurities (germanium) and pores are present in silicon.

\subsubsection{Two-Body Potential}

\begin{figure}[!h]
\includegraphics[width=8.5cm]{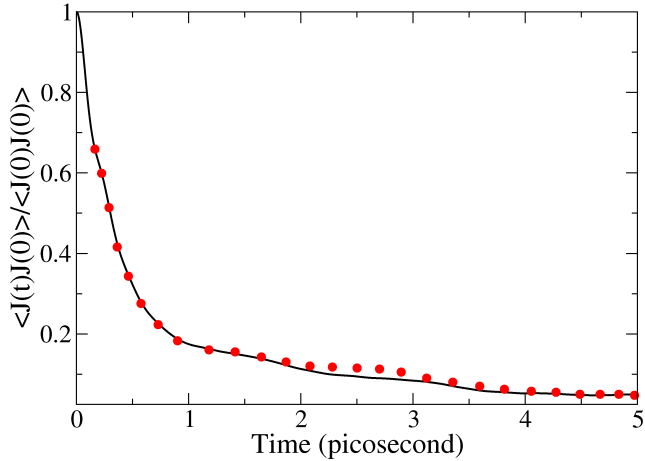}
\caption{\label{fig1}(Color online) Normalized HCACF obtained for 256 atom FCC argon in this work is represented by (black) solid line. The same function, calculated by McGaughey and Kaviany (Ref.~\onlinecite{McGaughey_2003}) is represented by (red) dotted line.}
\end{figure}

We calculate the lattice thermal conductivity of face centered cubic argon. We have chosen the same system size (256 atom) and equilibration procedure as in McGaughey and Kaviany~\cite{McGaughey_2003} in order to calculate HCACF. After equilibration, the system is allowed to run in NVE ensemble at 50 K for 2 ns with a time step of 0.1 femtosecond. We are aware   that this time step is small but we chose it to be consistent with the previous literature~\cite{McGaughey_2003} and to enable us to calculate more accurate numerical derivatives of $\bm(R)$. We performed five independent simulations to average HCACF's. The normalized HCACF is plotted in Fig.\ref{fig1} which is similar to what was obtained in McGaughey and Kaviany's work.

\begin{figure}[!h]
\includegraphics[width=8.5cm]{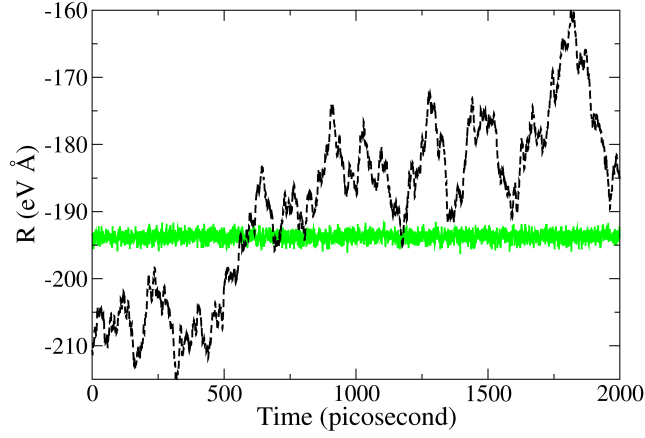}
\caption{\label{fig2}(Color online) Comparison of energy moments ($\bm{R}$) for \emph{formulation 1}, (black) dashed line, and \emph{formulation 2}, (green) solid line.}
\end{figure}

In Fig.~\ref{fig2} we have plotted the time evolution of $\bm{R}$. It is clear that there is a quantitative difference between \emph{formulation 1} and \emph{formulation 2}. The former, as expected, fluctuates around a constant value. Whereas the latter displays an increasing tendency.

\begin{figure}[!h]
\includegraphics[width=8.5cm]{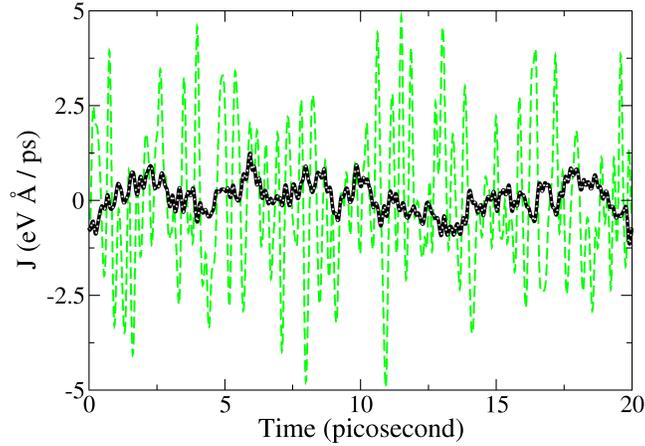}
\caption{\label{fig3}(Color online) Heat current obtained from analytical, (black) solid line, and numerical, (green) dashed line, derivatives of \emph{formulation 1}. (Grey) dotted line represents numerical derivative of \emph{formulation 2}.}
\end{figure}

In principle, the analytical and numerical derivatives of $\bm{R}$ should produce the same heat current ($\bm{J}$). Fig.~\ref{fig3} compares the behavior of the heat current obtained from the analytical and numerical derivative of $\bm{R}$. The inconsistency between the \emph{formulation 1} and analytical $\bm{J}$ is obvious. On the other hand, the numerical derivatives of \emph{formulation 2} perfectly matches the analytical form.

For the calculation of thermal conductivity, the system size is enlarged to 4000 atoms and the total simulation time is extended to 20 ns. Again, five different initial conditions are evaluated and the results are averaged. Ensemble average for the autocorrelation is obtained by selecting 6 millions time origins 3 fs apart from each other. In calculation of $\kappa$ from HCACF, the data was fitted to a double exponential, Eq.~\ref{eq:doublexp}, and the infinite integration was carried out over this function, an approach adopted by many previous studies~\cite{Che_2000_1,Che_2000_2,Schelling_2002,McGaughey_2003}.

\begin{equation}
f(t)=a_{1}e^{-t/\tau_{1}}+a_{2}e^{-t/\tau_{2}} \label{eq:doublexp}
\end{equation}

\begin{figure}[!h]
\includegraphics[width=8.5cm]{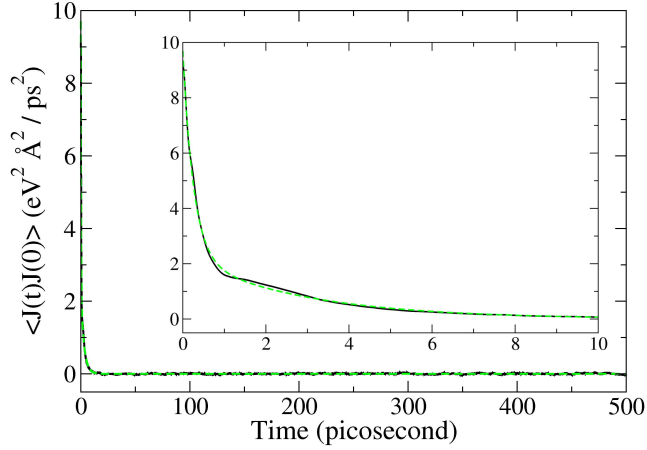}
\caption{\label{fig4}(Color online) Heat current autocorrelation function, (black) solid line, obtained for 4000 atom FCC argon at 50 K and the corresponding two exponential fit, (green) dashed line. The inset shows the same functions in the same units but for shorter correlation time i.e. 10 ps.}
\end{figure}

In this equation  $a_{1},a_{2},\tau_{1}$ and $\tau_{2}$ are parameters to be fitted and $t$ is the correlation time. Both the autocorrelation function and corresponding double exponential function are plotted in Fig.~\ref{fig4}. The thermal conductivity is calculated as 0.403$\pm$0.01 W/mK which is very close to the value (0.417 W/mK) given in Ref.~\onlinecite{McGaughey_2003}.

\begin{figure}[!h]
\includegraphics[width=8.5cm]{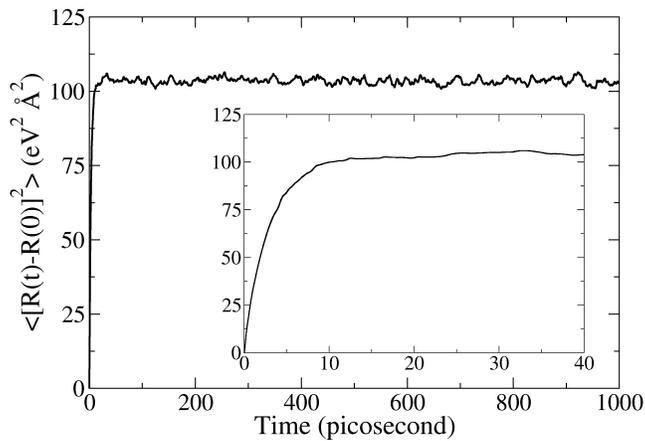}
\caption{\label{fig5} Einstein relation using \emph{formulation 1} obtained for 4000 atom FCC argon crystal at 50 K. The inlet shows the same function in the same units but for shorter time i.e. 40 ps.}
\end{figure}

The Einstein's relations corresponding to \emph{formulation 1} is presented in Fig.~\ref{fig5}. The formulation in Eq.~\ref{eq:einstein} shows that in order to obtain a thermal conductivity, there should be a finite positive slope of the given function. In the case of \emph{formulation 1}, two different behavior can be identified. First, there is a fast increase in the function and then a plateau is reached with no further change thereafter, see the inset of Fig.~\ref{fig5}. Due to this behavior, an ambiguity arises in the calculation of thermal conductivity. As time goes to infinity the total slope goes to zero giving a zero conductivity. Different numbers of time origins from 100 thousands to 6 millions have been tried to calculate ensemble average for Einstein's relation however the behavior of the curve did not change.

\begin{figure}[!h]
\includegraphics[width=8.5cm]{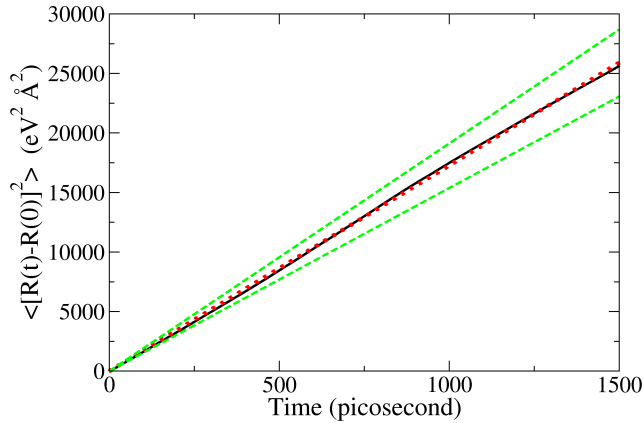}
\caption{\label{fig6}(Color online) Einstein relation using \emph{formulation 2}, (black) solid line, obtained for 4000 atom FCC argon crystal at 50 K and the corresponding error limits, (green) dashed lines. (Red) dotted line represents the linear fit Einstein relation.}
\end{figure}

The second model, on the other hand, has a finite positive slope as seen in Fig.~\ref{fig6}. This result has been obtained by using 200 thousands time origins with 90 fs apart.  In this figure, we also present the corresponding linear fit to Einstein relation and the upper-lower bounds (corresponding to five different initial conditions). The thermal conductivity from the linear fit is 0.416$\pm$0.05 W/mK which is very close to the value (0.403 W/mK) obtained by HCACF. The variance in the calculated values can be further reduced by using more time origins in a longer total run time.

In the previous section we pointed out that $bm{R}$ is composed of a kinetic ($\bm{R}_k$) and a potential ($\bm{R}_p$) portion and further stated that $\bm{R}_p$ should not contribute to the thermal conductivity of the non-convective solid system. For a diagonal element, Einstein relation can be separated and written as

\begin{align}
\lefteqn{\langle[R(t)-R(0)]^2\rangle =} & & \nonumber \\
& & \langle[R_k(t)-R_k(0)]^2\rangle+\langle[R_p(t)-R_p(0)]^2\rangle + \nonumber \\
& & 2\langle[R_p(t)-R_p(0)][R_k(t)-R_k(0)]\rangle
\label{eq:einstein_split}
\end{align}

\begin{figure}[!h]
\includegraphics[width=8.5cm]{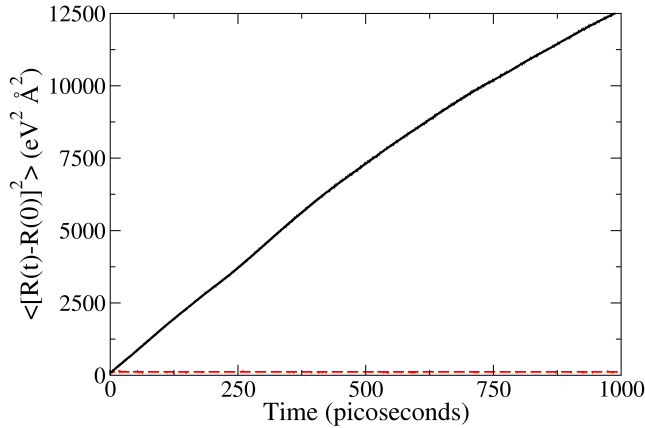}
\caption{\label{fig7}(Color online) Einstein relation for kinetic ($\bm{R}_k$), (black) solid line, and potential ($\bm{R}_p$), (red) dashed line, contributions to total $\bm{R}$ obtained for 4000 atom FCC argon crystal at 50 K.}
\end{figure}

We calculated the first and second terms on the right hand side of the Eq.~\ref{eq:einstein_split} and plotted in Fig.~\ref{fig7}. It can be seen that Einstein relation for $\bm{R}_k$ is equal to the Einstein relation for total $\bm{R}$. The second term on the right hand side of the Eq.~\ref{eq:einstein_split} is just a flat line and do not add to the total. Accordingly, the coupled term in Eq.~\ref{eq:einstein_split}, on the average, does not contribute to the thermal conductivity.

\subsubsection{N-Body Potential}

\begin{figure}[!h]
\includegraphics[width=8.5cm]{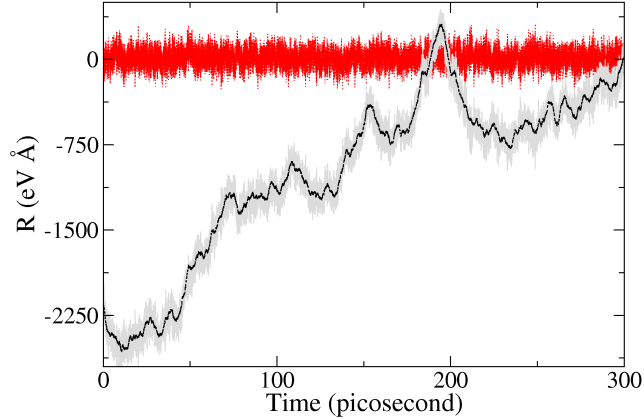}
\caption{\label{fig8}(Color online) Kinetic ($\bm{R}_k$), (black) dashed line, and potential ($\bm{R}_p$), (red) dotted line, portions of total energy moment, (grey) solid line. The data is obtained for 8000-atom silicon crystal at 300 K.}
\end{figure}

We selected silicon system represented by the Tersoff potential for many body simulations. We applied the same procedure to calculate Einstein relation for 10$\times$10$\times$10 supercell of Si, see Fig.~\ref{fig8}. By also simulating 6$\times$6$\times$6 and 8$\times$8$\times$8 we have established that the calculated conductivity is well converged within 5\%. The calculated thermal conductivity of silicon is 160.5$\pm$10.0 W/mK which is approximately 10\%  different than the experimental value of 142 - 148 W/mK ~\cite{Shanks_1963,Gesele_1997} at 300 K. Again, the Einstein relation for $\bm{R}_k$ is almost identical to the one for total $\bm{R}$. This confirms that the thermal conductivity does not depend on how the potential energy is split between interacting atoms and only the phonons are relevant to calculations. Ladd~\textit{et al.}~\cite{Ladd_1986} draw a similar conclusion that the heat current should be independent of the localization of the potential energy in a system at or near equilibrium. In their study, Schelling~\textit{et al.}~\cite{Schelling_2002} re-derived the heat current for different ways of energy distribution for a 3-body potential and calculated the thermal conductivity of silicon by using HCACF. Their results also did not differ for the reasons stated here. However, splitting may be important for the system where convective motion is dominant.

\subsection{Porosity and Ge-Clusters in Silicon}

\begin{figure}[!h]
\includegraphics[width=12.0cm]{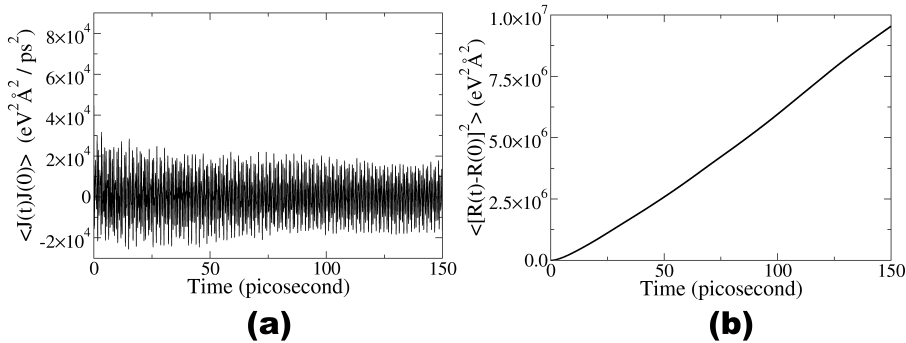}
\caption{\label{fig9}(Color online) (a) HCACF and (b) Einstein relation for silicon-germanium nano-composite having germanium clusters of size 1.65 nm. The clusters form a simple cube having a side length of 3.311 nm in a 12$\times$12$\times$12 supercell silicon matrix.}
\end{figure}

\begin{figure}[!h]
\includegraphics[width=8.5cm]{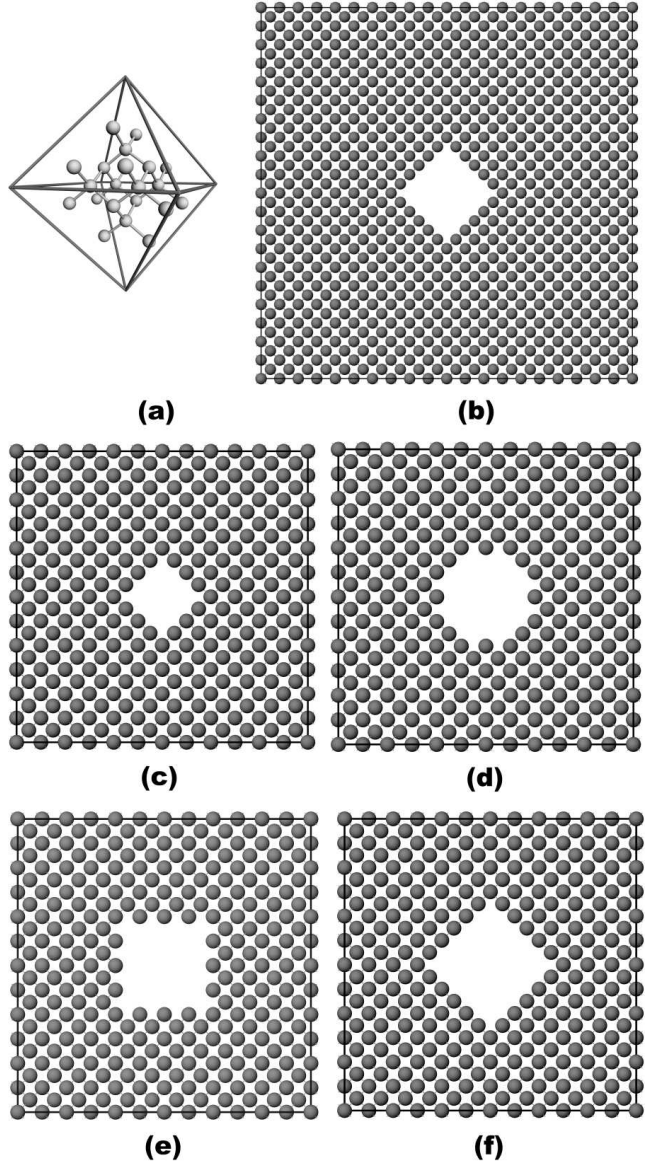}
\caption{\label{fig10} The atomistic arrangement of (a) 22-atom germanium clusters and (b)-(f) tubular vacancies in silicon having different size and shapes.}
\end{figure}

Heterogeneous systems contain a mixture of elements having different masses, elastic constants and lattice parameters. Due to this fact, HCACFs show high frequency fluctuations through its decay in heterogeneous materials. These high frequency fluctuations pose computational problems when thermal conductivity is calculated. In order to reveal the true decay time, HCACF must be filtered. This is generally done by transforming the function to frequency space and eliminating the high frequencies. Sometimes it is necessary to remove so many data that the back-transformed function becomes jagged and a good fit to a decaying function cannot be obtained. This problem is encountered very frequently because the heterogeneous materials are studied more often for their technological importance compared to their pure counterparts. Being heterogeneous systems, silicon-germanium alloys have the same problem when thermal conductivity is evaluated via HCACF. In order to point out the benefit of using Einstein relation in heterogeneous systems, we have calculated the thermal conductivity of silicon matrix which is embedded with Ge nano-clusters. Each cluster made up of 22 germanium atoms surrounded with a regular octahedron of silicon (approximately 1.65 nm in diameter), see Fig~\ref{fig10}(a). The surfaces of the octahedron consist of \{111\} family of planes. The clusters are placed in a 12$\times$12$\times$12 supercell of silicon and they form a periodic simple cubic arrangement having a side of 3.311 nm. In Fig~\ref{fig9}(a) we plotted the noisy HCACF obtained for the given nano-composite. Without filtering the high frequencies it is very difficult to fit this function to an exponential decay. On the other hand, in Fig.~\ref{fig9}(b) it can be seen that the Einstein relation behaves much better. The reason of this can be found in the integrated nature of Einstein relation where high frequency fluctuations in HCACF cancel out. The thermal conductivity is calculated as 21.74 W/mK for this system. This is an 80\% decrease over the value for pure silicon with only 1.3 at.\% Ge impurity. The clusters introduce internal surfaces where the opposite sides have different elastic constants, atomic masses and lattice parameters. Consequently, processes such as boundary and impurity scattering are heavily contributing to the thermal resistance of the lattice.

The minimum thermal conductivity in bulk silicon was estimated as 1.0-1.3 W/mK~\cite{Slack_1991,Allen_1989,Cahill_1988} when the structure is amorphous. These studies suggest that for amorphous silicon the phonon mean free path is on the order of average nearest neighbor distance. Later experimental studies claimed that even lower thermal conductivities than the ones in amorphous systems are possible in doped crystalline silicon with nanopores~\cite{Drost_1995,Gesele_1997,Lysenko_2000}. This modification, to lower extend (i.e. an order of magnitude decrease from bulk thermal conductivity) was also observed in microporous silicon by Song and Chen~\cite{Song_2004}. This result is not surprising because in addition to decrease in phonon mean free paths and relaxation times due to internal surface scattering, there are less conduction channels due to missing silicon atoms. We have tested these ideas on porosity with the corrected $\bm{R}$ and Einstein relation. The porous silicon structures and the corresponding thermal conductivities are given in Fig.~\ref{fig10} and Table~\ref{table1} respectively. In Table~\ref{table1}, a* represents the same octahedron in Fig.~\ref{fig10} without germanium atoms inside. Porosity is defined as the ratio of number of missing atoms to number of atoms in a perfect system. We have also presented the ratio of 2 bonded atoms to 4 bonded atoms (r2) and the ratio of 3 bonded atoms to 4 bonded atoms (r3) as an indication of the internal surface area. The tubular pores in Fig.~\ref{fig10}(b),(c) and (f) have \{110\} surface planes, Fig.~\ref{fig10}(d) has both \{100\} and \{110\} surface planes and Fig.~\ref{fig10}(e) has \{100\} surface planes.
In general, as the porosity and internal surface area increase, the thermal conductivity decreases as seen in Table~\ref{table1} in accordance with the previous findings from other theoretical methods ~\cite{Hopkins_2009,Alvarez_2010}. However, this trend breaks down for structures d, e and f. Although f has higher porosity and surface area than d and e, it has higher thermal conductivity. Structures d and e have atoms with only 2 bonds, they act as surface rattlers. These rattlers further decrease the conductivity. Among d and e, e has lower conductivity because it has higher porosity.

\begin{scriptsize}\begin{table}[!h]
\caption{\label{table1} Thermal conductivity (TC), porosity, the ratio of number of atoms with 2-bonds to number of atoms with 4-bonds, (r2) and number of atoms with 3-bonds to number of atoms with 4-bonds (r3) for the structures displayed in Fig~\ref{fig10}.}
\begin{ruledtabular}
\begin{tabular}{lccccc}
Struct.&TC (W/mK)&Porosity&r2&r3\\\hline
a*&31.58&0.0127&-&0.0143\\
b &23.55&0.0313&-&0.0265\\
c &15.11&0.0313&-&0.0449\\
d & 6.54&0.0729&0.0159&0.0478\\
e & 5.75&0.0868&0.0162&0.0486\\
f & 8.47&0.0868&-&0.0823\\\hline
\end{tabular}
\end{ruledtabular}
\end{table}\end{scriptsize}

\section{Concluding Remarks}

In summary, we have developed a correct definition for energy moment for systems evolving under classical dynamics. The thermal conductivity calculated from Einstein relation using the reformulated form of $\bm{R}$ yields the same result as the one calculated from HCACF, whereas the previous definition produces zero thermal conductivity in a non-diffusive solid system. We employed this method successfully for FCC argon and diamond silicon. Using $\bm{R}$ and associated Einstein relationship simplifies the calculations compared to HCACF's. It overcomes the difficulties inherent in taking the analytical derivative of $\bm{R}$ to obtain heat current. This also removes any ambiguity that arises from fitting the HCACF to a function in order to calculate thermal conductivity. Partition of R into potential and kinetic terms, moreover, demonstrates clearly the null contribution of potential energy term for any non-diffusive system.  However, this is not case for systems with diffusion (directly contributing mainly as the convective heat flow). We also have demonstrated the utility of the method for bulk Si structures.
With the studies performed on porous Si (cylindrical pores, and holes) and Ge-nanodots embedded into Si matrix, we have explored specific features of R-based or HCACF based thermal conductivity calculations.  In these systems, we observe increased noise on HCACF when compared to pure Si simulations, in contrast R-based calculation for these complex systems produced smoother behavior leading to well defined thermal conductivity values. In our earlier studies, we have extensively taken advantage of this formulation in determining thermal conductivity of Si, Si-Ge nanostructures,~\cite{JustinNanotech} graphene nanoribbons~\cite{JustinACS} and BN-nanostructures~\cite{Cem-BN-PRB}.

\begin{acknowledgments}

We acknowledge support from NSF (DMR 0844082) to the International Institute of Materials for Energy Conversion at Texas AM University as well as AFRL and DOE LLNS - INSER. Parts of the computations were carried out at the facilities of Laboratory of Computational Engineering of Nanomaterials, also supported by ARO and ONR grants. We also would like to thank the Supercomputing Center of Texas AM University for generous time allocation made for this project.

\end{acknowledgments}




\begin{thebibliography}{}%
\makeatletter
\providecommand \@ifxundefined [1]{%
 \ifx #1\undefined \expandafter \@firstoftwo
 \else \expandafter \@secondoftwo
\fi
}%
\providecommand \@ifnum [1]{%
 \ifnum #1\expandafter \@firstoftwo
 \else \expandafter \@secondoftwo
\fi
}%
\providecommand \enquote [1]{``#1''}%
\providecommand \bibnamefont  [1]{#1}%
\providecommand \bibfnamefont [1]{#1}%
\providecommand \citenamefont [1]{#1}%
\providecommand\href[0]{\@sanitize\@href}%
\providecommand\@href[1]{\endgroup\@@startlink{#1}\endgroup\@@href}%
\providecommand\@@href[1]{#1\@@endlink}%
\providecommand \@sanitize [0]{\begingroup\catcode`\&12\catcode`\#12\relax}%
\@ifxundefined \pdfoutput {\@firstoftwo}{%
 \@ifnum{\z@=\pdfoutput}{\@firstoftwo}{\@secondoftwo}%
}{%
 \providecommand\@@startlink[1]{\leavevmode\special{html:<a href="#1">}}%
 \providecommand\@@endlink[0]{\special{html:</a>}}%
}{%
 \providecommand\@@startlink[1]{%
  \leavevmode
  \pdfstartlink
   attr{/Border[0 0 1 ]/H/I/C[0 1 1]}%
   user{/Subtype/Link/A<</Type/Action/S/URI/URI(#1)>>}%
  \relax
 }%
 \providecommand\@@endlink[0]{\pdfendlink}%
}%
\providecommand \url  [0]{\begingroup\@sanitize \@url }%
\providecommand \@url [1]{\endgroup\@href {#1}{\urlprefix}}%
\providecommand \urlprefix [0]{URL }%
\providecommand \Eprint[0]{\href }%
\@ifxundefined \urlstyle {%
  \providecommand \doi [1]{doi:\discretionary{}{}{}#1}%
}{%
  \providecommand \doi [0]{doi:\discretionary{}{}{}\begingroup
  \urlstyle{rm}\Url }%
}%
\providecommand \doibase [0]{http://dx.doi.org/}%
\providecommand \Doi[1]{\href{\doibase#1}}%
\providecommand \selectlanguage [0]{\@gobble}%
\providecommand \bibinfo [0]{\@secondoftwo}%
\providecommand \bibfield [0]{\@secondoftwo}%
\providecommand \translation [1]{[#1]}%
\providecommand \BibitemOpen[0]{}%
\providecommand \bibitemStop [0]{}%
\providecommand \bibitemNoStop [0]{.\EOS\space}%
\providecommand \EOS [0]{\spacefactor3000\relax}%
\providecommand \BibitemShut [1]{\csname bibitem#1\endcsname}%
\end{thebibliography}%


\begin{thebibliography}{10}%
\makeatletter
\providecommand \@ifxundefined [1]{%
 \ifx #1\undefined \expandafter \@firstoftwo
 \else \expandafter \@secondoftwo
\fi
}%
\providecommand \@ifnum [1]{%
 \ifnum #1\expandafter \@firstoftwo
 \else \expandafter \@secondoftwo
\fi
}%
\providecommand \enquote [1]{``#1''}%
\providecommand \bibnamefont  [1]{#1}%
\providecommand \bibfnamefont [1]{#1}%
\providecommand \citenamefont [1]{#1}%
\providecommand\href[0]{\@sanitize\@href}%
\providecommand\@href[1]{\endgroup\@@startlink{#1}\endgroup\@@href}%
\providecommand\@@href[1]{#1\@@endlink}%
\providecommand \@sanitize [0]{\begingroup\catcode`\&12\catcode`\#12\relax}%
\@ifxundefined \pdfoutput {\@firstoftwo}{%
 \@ifnum{\z@=\pdfoutput}{\@firstoftwo}{\@secondoftwo}%
}{%
 \providecommand\@@startlink[1]{\leavevmode}%
 \providecommand\@@endlink[0]{}%
}{%
 \providecommand\@@startlink[1]{%
  \leavevmode
  \pdfstartlink
   attr{/Border[0 0 1 ]/H/I/C[0 1 1]}%
   user{/Subtype/Link/A<</Type/Action/S/URI/URI(#1)>>}%
  \relax
 }%
 \providecommand\@@endlink[0]{\pdfendlink}%
}%
\providecommand \url  [0]{\begingroup\@sanitize \@url }%
\providecommand \@url [1]{\endgroup\@href {#1}{\urlprefix}}%
\providecommand \urlprefix [0]{URL }%
\providecommand \Eprint[0]{\href }%
\@ifxundefined \urlstyle {%
  \providecommand \doi [1]{doi:\discretionary{}{}{}#1}%
}{%
  \providecommand \doi [0]{doi:\discretionary{}{}{}\begingroup
  \urlstyle{rm}\Url }%
}%
\providecommand \doibase [0]{http://dx.doi.org/}%
\providecommand \Doi[1]{\href{\doibase#1}}%
\providecommand \selectlanguage [0]{\@gobble}%
\providecommand \bibinfo [0]{\@secondoftwo}%
\providecommand \bibfield [0]{\@secondoftwo}%
\providecommand \translation [1]{[#1]}%
\providecommand \BibitemOpen[0]{}%
\providecommand \bibitemStop [0]{}%
\providecommand \bibitemNoStop [0]{.\EOS\space}%
\providecommand \EOS [0]{\spacefactor3000\relax}%
\providecommand \BibitemShut [1]{\csname bibitem#1\endcsname}%
\bibitem{Rowe_2006}%
  \BibitemOpen
  \emph{\bibinfo {title} {Thermoelectrics handbook, macro to nano}},\ edited
  by\ \bibinfo {editor} {\bibfnamefont{D.~M.}\ \bibnamefont{Rowe}}\ (\bibinfo
  {publisher} {CRC Press-Taylor and Francis Group, Boca Raton},\ \bibinfo
  {year} {2006})\BibitemShut{NoStop}%
\bibitem{Snyder_2008}%
  \BibitemOpen
  \bibfield{author}{%
  \bibinfo {author} {\bibfnamefont{G.~J.}\ \bibnamefont{Snyder}}\ and\ \bibinfo
  {author} {\bibfnamefont{E.~S.}\ \bibnamefont{Toberer}},\ }%
  \bibfield{journal}{%
  \bibinfo {journal} {Nat. Mater.}\ }%
  \textbf{\bibinfo {volume} {7}},\ \bibinfo {pages} {105} (\bibinfo {year}
  {2008})\BibitemShut{NoStop}%
\bibitem{Dames_2004}%
  \BibitemOpen
  \bibfield{author}{%
  \bibinfo {author} {\bibfnamefont{C.}~\bibnamefont{Dames}}\ and\ \bibinfo
  {author} {\bibfnamefont{G.}~\bibnamefont{Chen}},\ }%
  \bibfield{journal}{%
  \bibinfo {journal} {J. Appl. Phys.}\ }%
  \textbf{\bibinfo {volume} {95}},\ \bibinfo {pages} {682} (\bibinfo {year}
  {2004})\BibitemShut{NoStop}%
\bibitem{Hochbaum_2008}%
  \BibitemOpen
  \bibfield{author}{%
  \bibinfo {author} {\bibfnamefont{A.~I.}\ \bibnamefont{Hochbaum}}, \bibinfo
  {author} {\bibfnamefont{R.}~\bibnamefont{Chen}}, \bibinfo {author}
  {\bibfnamefont{R.~D.}\ \bibnamefont{Delgado}}, \bibinfo {author}
  {\bibfnamefont{W.}~\bibnamefont{Liang}}, \bibinfo {author}
  {\bibfnamefont{E.~C.}\ \bibnamefont{Garnett}}, \bibinfo {author}
  {\bibfnamefont{M.}~\bibnamefont{Najarian}}, \bibinfo {author}
  {\bibfnamefont{A.}~\bibnamefont{Majumdar}},\ and\ \bibinfo {author}
  {\bibfnamefont{P.}~\bibnamefont{Yang}},\ }%
  \bibfield{journal}{%
  \bibinfo {journal} {Nature}\ }%
  \textbf{\bibinfo {volume} {451}},\ \bibinfo {pages} {163} (\bibinfo {year}
  {2008})\BibitemShut{NoStop}%
\bibitem{Balandin_1998}%
  \BibitemOpen
  \bibfield{author}{%
  \bibinfo {author} {\bibfnamefont{A.}~\bibnamefont{Balandin}}\ and\ \bibinfo
  {author} {\bibfnamefont{K.~L.}\ \bibnamefont{Wang}},\ }%
  \bibfield{journal}{%
  \bibinfo {journal} {Phys. Rev. B}\ }%
  \textbf{\bibinfo {volume} {58}},\ \bibinfo {pages} {1544} (\bibinfo {year}
  {1998})\BibitemShut{NoStop}%
\bibitem{Chen_2001_1}%
  \BibitemOpen
  \bibfield{author}{%
  \bibinfo {author} {\bibfnamefont{G.}~\bibnamefont{Chen}},\ }%
  \bibfield{journal}{%
  \bibinfo {journal} {Phys. Rev. Lett.}\ }%
  \textbf{\bibinfo {volume} {86}},\ \bibinfo {pages} {2297} (\bibinfo {year}
  {2001})\BibitemShut{NoStop}%
\bibitem{Chen_2001_2}%
  \BibitemOpen
  \bibfield{author}{%
  \bibinfo {author} {\bibfnamefont{G.}~\bibnamefont{Chen}}\ and\ \bibinfo
  {author} {\bibfnamefont{T.}~\bibnamefont{Zeng}},\ }%
  \bibfield{journal}{%
  \bibinfo {journal} {Microscale Term. Eng.}\ }%
  \textbf{\bibinfo {volume} {5}},\ \bibinfo {pages} {71} (\bibinfo {year}
  {2001})\BibitemShut{NoStop}%
\bibitem{Majumdar_1993}%
  \BibitemOpen
  \bibfield{author}{%
  \bibinfo {author} {\bibfnamefont{A.}~\bibnamefont{Majumdar}},\ }%
  \bibfield{journal}{%
  \bibinfo {journal} {J. Heat Transf.}\ }%
  \textbf{\bibinfo {volume} {115}},\ \bibinfo {pages} {7} (\bibinfo {year}
  {1993})\BibitemShut{NoStop}%
\bibitem{Turney_2009}%
  \BibitemOpen
  \bibfield{author}{%
  \bibinfo {author} {\bibfnamefont{J.~E.}\ \bibnamefont{Turney}}, \bibinfo
  {author} {\bibfnamefont{E.~S.}\ \bibnamefont{Landry}}, \bibinfo {author}
  {\bibfnamefont{A.~J.~H.}\ \bibnamefont{McGaughey}},\ and\ \bibinfo {author}
  {\bibfnamefont{C.~H.}\ \bibnamefont{Amon}},\ }%
  \bibfield{journal}{%
  \bibinfo {journal} {Phys. Rev. B}\ }%
  \textbf{\bibinfo {volume} {79}},\ \bibinfo {pages} {064301} (\bibinfo {year}
  {2009})\BibitemShut{NoStop}%
\bibitem{Zhao_2008}%
  \BibitemOpen
  \bibfield{author}{%
  \bibinfo {author} {\bibfnamefont{H.}~\bibnamefont{Zhao}}\ and\ \bibinfo
  {author} {\bibfnamefont{J.~B.}\ \bibnamefont{Freund}},\ }%
  \bibfield{journal}{%
  \bibinfo {journal} {J. Appl. Phys.}\ }%
  \textbf{\bibinfo {volume} {104}},\ \bibinfo {pages} {033514} (\bibinfo {year}
  {2008})\BibitemShut{NoStop}%
\bibitem{Henry_2008}%
  \BibitemOpen
  \bibfield{author}{%
  \bibinfo {author} {\bibfnamefont{A.}~\bibnamefont{Henry}}\ and\ \bibinfo
  {author} {\bibfnamefont{G.}~\bibnamefont{Chen}},\ }%
  \bibfield{journal}{%
  \bibinfo {journal} {J. Comput. Theor. Nanosci.}\ }%
  \textbf{\bibinfo {volume} {5}},\ \bibinfo {pages} {141} (\bibinfo {year}
  {2008})\BibitemShut{NoStop}%
\bibitem{Volz_2007}%
  \BibitemOpen
  \emph{\bibinfo {title} {Topics in Applied Physics : Microscale and Nanoscale
  Heat Transfer}},\ edited by\ \bibinfo {editor}
  {\bibfnamefont{S.}~\bibnamefont{Volz}},\ Vol.\ \bibinfo {volume} {107}\
  (\bibinfo {publisher} {Springer},\ \bibinfo {address} {New York},\ \bibinfo
  {year} {2007})\BibitemShut{NoStop}%
\bibitem{Cahill_2003}%
  \BibitemOpen
  \bibfield{author}{%
  \bibinfo {author} {\bibfnamefont{D.~G.}\ \bibnamefont{Cahill}}, \bibinfo
  {author} {\bibfnamefont{W.~K.}\ \bibnamefont{Ford}}, \bibinfo {author}
  {\bibfnamefont{K.~E.}\ \bibnamefont{Goodson}}, \bibinfo {author}
  {\bibfnamefont{G.~D.}\ \bibnamefont{Mahan}}, \bibinfo {author}
  {\bibfnamefont{A.}~\bibnamefont{Majumdar}}, \bibinfo {author}
  {\bibfnamefont{H.~J.}\ \bibnamefont{Maris}}, \bibinfo {author}
  {\bibfnamefont{R.}~\bibnamefont{Merlin}},\ and\ \bibinfo {author}
  {\bibfnamefont{S.~R.}\ \bibnamefont{Phillpot}},\ }%
  \bibfield{journal}{%
  \bibinfo {journal} {J. Appl. Phys.}\ }%
  \textbf{\bibinfo {volume} {93}},\ \bibinfo {pages} {793} (\bibinfo {year}
  {2003})\BibitemShut{NoStop}%
\bibitem{Che_2000_1}%
  \BibitemOpen
  \bibfield{author}{%
  \bibinfo {author} {\bibfnamefont{J.}~\bibnamefont{Che}}, \bibinfo {author}
  {\bibfnamefont{T.}~\bibnamefont{{\c C}a{\u g}{\i}n}}, \bibinfo {author}
  {\bibfnamefont{W.}~\bibnamefont{Deng}},\ and\ \bibinfo {author}
  {\bibfnamefont{W.~A.}\ \bibnamefont{Goddard~III}},\ }%
  \bibfield{journal}{%
  \bibinfo {journal} {J. Chem. Phys.}\ }%
  \textbf{\bibinfo {volume} {113}},\ \bibinfo {pages} {6888} (\bibinfo {year}
  {2000})\BibitemShut{NoStop}%
\bibitem{Egelstaff_1962}%
  \BibitemOpen
  \bibfield{author}{%
  \bibinfo {author} {\bibfnamefont{P.~A.}\ \bibnamefont{Egelstaff}},\ }%
  \bibfield{journal}{%
  \bibinfo {journal} {Adv. Phys.}\ }%
  \textbf{\bibinfo {volume} {11}},\ \bibinfo {pages} {203} (\bibinfo {year}
  {1962})\BibitemShut{NoStop}%
\bibitem{Lukes_2000}%
  \BibitemOpen
  \bibfield{author}{%
  \bibinfo {author} {\bibfnamefont{J.~R.}\ \bibnamefont{Lukes}}, \bibinfo
  {author} {\bibfnamefont{D.~Y.}\ \bibnamefont{Li}}, \bibinfo {author}
  {\bibfnamefont{X.~G.}\ \bibnamefont{Liang}},\ and\ \bibinfo {author}
  {\bibfnamefont{C.~L.}\ \bibnamefont{Tien}},\ }%
  \bibfield{journal}{%
  \bibinfo {journal} {J. Heat Transf.}\ }%
  \textbf{\bibinfo {volume} {122}},\ \bibinfo {pages} {536} (\bibinfo {year}
  {2000})\BibitemShut{NoStop}%
\bibitem{Volz_1996}%
  \BibitemOpen
  \bibfield{author}{%
  \bibinfo {author} {\bibfnamefont{S.}~\bibnamefont{Volz}}, \bibinfo {author}
  {\bibfnamefont{J.~B.}\ \bibnamefont{Saulnier}}, \bibinfo {author}
  {\bibfnamefont{M.}~\bibnamefont{Lallemand}}, \bibinfo {author}
  {\bibfnamefont{B.}~\bibnamefont{Perrin}}, \bibinfo {author}
  {\bibfnamefont{P.}~\bibnamefont{Depondt}},\ and\ \bibinfo {author}
  {\bibfnamefont{M.}~\bibnamefont{Mareschal}},\ }%
  \bibfield{journal}{%
  \bibinfo {journal} {Phys. Rev. B}\ }%
  \textbf{\bibinfo {volume} {54}},\ \bibinfo {pages} {340} (\bibinfo {year}
  {1996})\BibitemShut{NoStop}%
\bibitem{Chen_2004}%
  \BibitemOpen
  \bibfield{author}{%
  \bibinfo {author} {\bibfnamefont{Y.}~\bibnamefont{Chen}}, \bibinfo {author}
  {\bibfnamefont{D.}~\bibnamefont{Li}}, \bibinfo {author}
  {\bibfnamefont{J.}~\bibnamefont{Yang}}, \bibinfo {author}
  {\bibfnamefont{Y.}~\bibnamefont{Wu}}, \bibinfo {author}
  {\bibfnamefont{J.~R.}\ \bibnamefont{Lukes}},\ and\ \bibinfo {author}
  {\bibfnamefont{A.}~\bibnamefont{Majumdar}},\ }%
  \bibfield{journal}{%
  \bibinfo {journal} {Physica B}\ }%
  \textbf{\bibinfo {volume} {394}},\ \bibinfo {pages} {270} (\bibinfo {year}
  {2004})\BibitemShut{NoStop}%
\bibitem{Maiti_1997}%
  \BibitemOpen
  \bibfield{author}{%
  \bibinfo {author} {\bibfnamefont{A.}~\bibnamefont{Maiti}}, \bibinfo {author}
  {\bibfnamefont{G.~D.}\ \bibnamefont{Mahan}},\ and\ \bibinfo {author}
  {\bibfnamefont{S.~T.}\ \bibnamefont{Pantelides}},\ }%
  \bibfield{journal}{%
  \bibinfo {journal} {Solid State Commun.}\ }%
  \textbf{\bibinfo {volume} {102}},\ \bibinfo {pages} {517} (\bibinfo {year}
  {1997})\BibitemShut{NoStop}%
\bibitem{Padgett_2006}%
  \BibitemOpen
  \bibfield{author}{%
  \bibinfo {author} {\bibfnamefont{C.~W.}\ \bibnamefont{Padgett}}, \bibinfo
  {author} {\bibfnamefont{O.}~\bibnamefont{Shenderova}},\ and\ \bibinfo
  {author} {\bibfnamefont{D.~W.}\ \bibnamefont{Brenner}},\ }%
  \bibfield{journal}{%
  \bibinfo {journal} {Nano Lett.}\ }%
  \textbf{\bibinfo {volume} {6}},\ \bibinfo {pages} {1827} (\bibinfo {year}
  {2006})\BibitemShut{NoStop}%
\bibitem{Cem-NL}%
  \BibitemOpen
  \bibfield{author}{%
  \bibinfo {author} {\bibfnamefont{C.}~\bibnamefont{Sevik}}, \bibinfo {author}
  {\bibfnamefont{H.}~\bibnamefont{Sevin\ifmmode\mbox{\c{c}}\else\c{c}\fi{}li}},
  \bibinfo {author} {\bibfnamefont{G.}~\bibnamefont{Cuniberti}},\ and\ \bibinfo
  {author} {\bibfnamefont{T.}~\bibnamefont{\c{C}a\u{g}{\i}n}},\ }%
  \bibfield{journal}{%
  \bibinfo {journal} {Nano Letters}\ }%
  \textbf{\bibinfo {volume} {11}},\ \bibinfo {pages} {4971} (\bibinfo {year}
  {2011})\BibitemShut{NoStop}%
\bibitem{Evans_1982}%
  \BibitemOpen
  \bibfield{author}{%
  \bibinfo {author} {\bibfnamefont{D.~J.}\ \bibnamefont{Evans}},\ }%
  \bibfield{journal}{%
  \bibinfo {journal} {Phys. Lett. A}\ }%
  \textbf{\bibinfo {volume} {91}},\ \bibinfo {pages} {457} (\bibinfo {year}
  {1982})\BibitemShut{NoStop}%
\bibitem{Green_1954}%
  \BibitemOpen
  \bibfield{author}{%
  \bibinfo {author} {\bibfnamefont{M.~S.}\ \bibnamefont{Green}},\ }%
  \bibfield{journal}{%
  \bibinfo {journal} {J. Chem. Phys.}\ }%
  \textbf{\bibinfo {volume} {22}},\ \bibinfo {pages} {398} (\bibinfo {year}
  {1954})\BibitemShut{NoStop}%
\bibitem{Kubo_1957}%
  \BibitemOpen
  \bibfield{author}{%
  \bibinfo {author} {\bibfnamefont{R.}~\bibnamefont{Kubo}},\ }%
  \bibfield{journal}{%
  \bibinfo {journal} {J. Phys. Soc. Japan}\ }%
  \textbf{\bibinfo {volume} {12}},\ \bibinfo {pages} {570} (\bibinfo {year}
  {1957})\BibitemShut{NoStop}%
\bibitem{Zwanzig_1965}%
  \BibitemOpen
  \bibfield{author}{%
  \bibinfo {author} {\bibfnamefont{R.}~\bibnamefont{Zwanzig}},\ }%
  \bibfield{journal}{%
  \bibinfo {journal} {Annu. Rev. Phys. Chem.}\ }%
  \textbf{\bibinfo {volume} {16}},\ \bibinfo {pages} {67} (\bibinfo {year}
  {1965})\BibitemShut{NoStop}%
\bibitem{McQuarrie_2000}%
  \BibitemOpen
  \bibfield{author}{%
  \bibinfo {author} {\bibfnamefont{D.~A.}\ \bibnamefont{McQuarrie}},\ }%
  \emph{\bibinfo {title} {Statistical Mechanics}}\ (\bibinfo {publisher}
  {University Science Books},\ \bibinfo {address} {California},\ \bibinfo
  {year} {2000})\BibitemShut{NoStop}%
\bibitem{Allen_1987}%
  \BibitemOpen
  \bibfield{author}{%
  \bibinfo {author} {\bibfnamefont{M.~P.}\ \bibnamefont{Allen}}\ and\ \bibinfo
  {author} {\bibfnamefont{D.~J.}\ \bibnamefont{Tildesley}},\ }%
  \emph{\bibinfo {title} {Computer Simulation of Liquids}}\ (\bibinfo
  {publisher} {Oxford University Press},\ \bibinfo {address} {Oxford},\
  \bibinfo {year} {1987})\BibitemShut{NoStop}%
\bibitem{Schelling_2002}%
  \BibitemOpen
  \bibfield{author}{%
  \bibinfo {author} {\bibfnamefont{P.~K.}\ \bibnamefont{Schelling}}, \bibinfo
  {author} {\bibfnamefont{S.~R.}\ \bibnamefont{Phillpot}},\ and\ \bibinfo
  {author} {\bibfnamefont{P.}~\bibnamefont{Keblinski}},\ }%
  \bibfield{journal}{%
  \bibinfo {journal} {Phys. Rev. B}\ }%
  \textbf{\bibinfo {volume} {65}},\ \bibinfo {pages} {144306} (\bibinfo {year}
  {2002})\BibitemShut{NoStop}%
\bibitem{Huang_2007}%
  \BibitemOpen
  \bibfield{author}{%
  \bibinfo {author} {\bibfnamefont{B.~L.}\ \bibnamefont{Huang}}, \bibinfo
  {author} {\bibfnamefont{A.~J.~H.}\ \bibnamefont{McGaughey}},\ and\ \bibinfo
  {author} {\bibfnamefont{M.}~\bibnamefont{Kaviany}},\ }%
  \bibfield{journal}{%
  \bibinfo {journal} {Int. J. Heat Mass Tran.}\ }%
  \textbf{\bibinfo {volume} {50}},\ \bibinfo {pages} {393} (\bibinfo {year}
  {2007})\BibitemShut{NoStop}%
\bibitem{Che_2000_2}%
  \BibitemOpen
  \bibfield{author}{%
  \bibinfo {author} {\bibfnamefont{J.}~\bibnamefont{Che}}, \bibinfo {author}
  {\bibfnamefont{T.}~\bibnamefont{{\c C}a{\u g}{\i}n}},\ and\ \bibinfo {author}
  {\bibfnamefont{W.~A.}\ \bibnamefont{Goddard~III}},\ }%
  \bibfield{journal}{%
  \bibinfo {journal} {Nanotechnology}\ }%
  \textbf{\bibinfo {volume} {11}},\ \bibinfo {pages} {65} (\bibinfo {year}
  {2000})\BibitemShut{NoStop}%
\bibitem{Evans_2010}%
  \BibitemOpen
  \bibfield{author}{%
  \bibinfo {author} {\bibfnamefont{W.~J.}\ \bibnamefont{Evans}}, \bibinfo
  {author} {\bibfnamefont{L.}~\bibnamefont{Hu}},\ and\ \bibinfo {author}
  {\bibfnamefont{P.}~\bibnamefont{Keblinski}},\ }%
  \bibfield{journal}{%
  \bibinfo {journal} {Appl. Phys. Lett.}\ }%
  \textbf{\bibinfo {volume} {96}},\ \bibinfo {pages} {203112} (\bibinfo {year}
  {2010})\BibitemShut{NoStop}%
\bibitem{Domingues_2005}%
  \BibitemOpen
  \bibfield{author}{%
  \bibinfo {author} {\bibfnamefont{G.}~\bibnamefont{Domingues}}, \bibinfo
  {author} {\bibfnamefont{S.}~\bibnamefont{Volz}}, \bibinfo {author}
  {\bibfnamefont{K.}~\bibnamefont{Joulain}},\ and\ \bibinfo {author}
  {\bibfnamefont{J.~J.}\ \bibnamefont{Greffet}},\ }%
  \bibfield{journal}{%
  \bibinfo {journal} {Phys. Rev. Lett.}\ }%
  \textbf{\bibinfo {volume} {94}},\ \bibinfo {pages} {085901} (\bibinfo {year}
  {2005})\BibitemShut{NoStop}%
\bibitem{Allen_1993}%
  \BibitemOpen
  \bibfield{author}{%
  \bibinfo {author} {\bibfnamefont{M.~P.}\ \bibnamefont{Allen}}\ and\ \bibinfo
  {author} {\bibfnamefont{A.~J.}\ \bibnamefont{Masters}},\ }%
  \bibfield{journal}{%
  \bibinfo {journal} {Mol. Phys.}\ }%
  \textbf{\bibinfo {volume} {79}},\ \bibinfo {pages} {435} (\bibinfo {year}
  {1993})\BibitemShut{NoStop}%
\bibitem{Rapaport_2004}%
  \BibitemOpen
  \bibfield{author}{%
  \bibinfo {author} {\bibfnamefont{D.~C.}\ \bibnamefont{Rapaport}},\ }%
  \emph{\bibinfo {title} {The Art of Molecular Dynamics Simulation}}\ (\bibinfo
  {publisher} {Cambridge University Press},\ \bibinfo {address} {Cambridge},\
  \bibinfo {year} {2004})\BibitemShut{NoStop}%
\bibitem{Tersoff_1989}%
  \BibitemOpen
  \bibfield{author}{%
  \bibinfo {author} {\bibfnamefont{J.}~\bibnamefont{Tersoff}},\ }%
  \bibfield{journal}{%
  \bibinfo {journal} {Phys. Rev. B}\ }%
  \textbf{\bibinfo {volume} {39}},\ \bibinfo {pages} {5566} (\bibinfo {year}
  {1989})\BibitemShut{NoStop}%
\bibitem{Donadio_2010}%
  \BibitemOpen
  \bibfield{author}{%
  \bibinfo {author} {\bibfnamefont{D.}~\bibnamefont{Donadio}}\ and\ \bibinfo
  {author} {\bibfnamefont{G.}~\bibnamefont{Galli}},\ }%
  \bibfield{journal}{%
  \bibinfo {journal} {Nano Lett.}\ }%
  \textbf{\bibinfo {volume} {10}},\ \bibinfo {pages} {847} (\bibinfo {year}
  {2010})\BibitemShut{NoStop}%
\bibitem{McGaughey_2003}%
  \BibitemOpen
  \bibfield{author}{%
  \bibinfo {author} {\bibfnamefont{A.~J.~H.}\ \bibnamefont{McGaughey}}\ and\
  \bibinfo {author} {\bibfnamefont{M.}~\bibnamefont{Kaviany}},\ }%
  \bibfield{journal}{%
  \bibinfo {journal} {Int. J. Heat Mass Tran.}\ }%
  \textbf{\bibinfo {volume} {47}},\ \bibinfo {pages} {1783} (\bibinfo {year}
  {2004})\BibitemShut{NoStop}%
\bibitem{Shanks_1963}%
  \BibitemOpen
  \bibfield{author}{%
  \bibinfo {author} {\bibfnamefont{H.~R.}\ \bibnamefont{Shanks}}, \bibinfo
  {author} {\bibfnamefont{P.~D.}\ \bibnamefont{Maycock}}, \bibinfo {author}
  {\bibfnamefont{P.}~\bibnamefont{Sidles}},\ and\ \bibinfo {author}
  {\bibfnamefont{G.~C.}\ \bibnamefont{Danielson}},\ }%
  \bibfield{journal}{%
  \bibinfo {journal} {Phys. Rev.}\ }%
  \textbf{\bibinfo {volume} {130}},\ \bibinfo {pages} {1743} (\bibinfo {year}
  {1963})\BibitemShut{NoStop}%
\bibitem{Gesele_1997}%
  \BibitemOpen
  \bibfield{author}{%
  \bibinfo {author} {\bibfnamefont{G.}~\bibnamefont{Gesele}}, \bibinfo {author}
  {\bibfnamefont{J.}~\bibnamefont{Linsmeier}}, \bibinfo {author}
  {\bibfnamefont{V.}~\bibnamefont{Drach}}, \bibinfo {author}
  {\bibfnamefont{J.}~\bibnamefont{Fricke}},\ and\ \bibinfo {author}
  {\bibfnamefont{R.}~\bibnamefont{Arens-Fischer}},\ }%
  \bibfield{journal}{%
  \bibinfo {journal} {J. Phys. D: Appl. Phys.}\ }%
  \textbf{\bibinfo {volume} {30}},\ \bibinfo {pages} {2911} (\bibinfo {year}
  {1997})\BibitemShut{NoStop}%
\bibitem{Ladd_1986}%
  \BibitemOpen
  \bibfield{author}{%
  \bibinfo {author} {\bibfnamefont{A.~J.~C.}\ \bibnamefont{Ladd}}, \bibinfo
  {author} {\bibfnamefont{B.}~\bibnamefont{Moran}},\ and\ \bibinfo {author}
  {\bibfnamefont{W.~G.}\ \bibnamefont{Hoover}},\ }%
  \bibfield{journal}{%
  \bibinfo {journal} {Phys. Rev. B}\ }%
  \textbf{\bibinfo {volume} {34}},\ \bibinfo {pages} {5058} (\bibinfo {year}
  {1986})\BibitemShut{NoStop}%
\bibitem{Slack_1991}%
  \BibitemOpen
  \bibfield{author}{%
  \bibinfo {author} {\bibfnamefont{G.~A.}\ \bibnamefont{Slack}}\ and\ \bibinfo
  {author} {\bibfnamefont{M.~A.}\ \bibnamefont{Hussain}},\ }%
  \bibfield{journal}{%
  \bibinfo {journal} {J. Appl. Phys.}\ }%
  \textbf{\bibinfo {volume} {70}},\ \bibinfo {pages} {2694} (\bibinfo {year}
  {1991})\BibitemShut{NoStop}%
\bibitem{Allen_1989}%
  \BibitemOpen
  \bibfield{author}{%
  \bibinfo {author} {\bibfnamefont{P.~B.}\ \bibnamefont{Allen}}\ and\ \bibinfo
  {author} {\bibfnamefont{J.~L.}\ \bibnamefont{Feldman}},\ }%
  \bibfield{journal}{%
  \bibinfo {journal} {Phys. Rev. Lett.}\ }%
  \textbf{\bibinfo {volume} {62}},\ \bibinfo {pages} {645} (\bibinfo {year}
  {1989})\BibitemShut{NoStop}%
\bibitem{Cahill_1988}%
  \BibitemOpen
  \bibfield{author}{%
  \bibinfo {author} {\bibfnamefont{D.~G.}\ \bibnamefont{Cahill}}, \bibinfo
  {author} {\bibfnamefont{H.~E.}\ \bibnamefont{Fischer}}, \bibinfo {author}
  {\bibfnamefont{T.}~\bibnamefont{Klitsner}}, \bibinfo {author}
  {\bibfnamefont{E.~T.}\ \bibnamefont{Swartz}},\ and\ \bibinfo {author}
  {\bibfnamefont{R.~O.}\ \bibnamefont{Pohl}},\ }%
  \bibfield{journal}{%
  \bibinfo {journal} {J. Vac. Sci. Technol. A}\ }%
  \textbf{\bibinfo {volume} {7}},\ \bibinfo {pages} {1259} (\bibinfo {year}
  {1988})\BibitemShut{NoStop}%
\bibitem{Drost_1995}%
  \BibitemOpen
  \bibfield{author}{%
  \bibinfo {author} {\bibfnamefont{A.}~\bibnamefont{Drost}}, \bibinfo {author}
  {\bibfnamefont{P.}~\bibnamefont{Steiner}}, \bibinfo {author}
  {\bibfnamefont{H.}~\bibnamefont{Moser}},\ and\ \bibinfo {author}
  {\bibfnamefont{W.}~\bibnamefont{Lang}},\ }%
  \bibfield{journal}{%
  \bibinfo {journal} {Sensor. Mater.}\ }%
  \textbf{\bibinfo {volume} {7}},\ \bibinfo {pages} {111} (\bibinfo {year}
  {1995})\BibitemShut{NoStop}%
\bibitem{Lysenko_2000}%
  \BibitemOpen
  \bibfield{author}{%
  \bibinfo {author} {\bibfnamefont{V.}~\bibnamefont{Lysenko}}, \bibinfo
  {author} {\bibfnamefont{B.}~\bibnamefont{Remaki}},\ and\ \bibinfo {author}
  {\bibfnamefont{D.}~\bibnamefont{Barbier}},\ }%
  \bibfield{journal}{%
  \bibinfo {journal} {Adv. Mater.}\ }%
  \textbf{\bibinfo {volume} {12}},\ \bibinfo {pages} {516} (\bibinfo {year}
  {2000})\BibitemShut{NoStop}%
\bibitem{Song_2004}%
  \BibitemOpen
  \bibfield{author}{%
  \bibinfo {author} {\bibfnamefont{D.}~\bibnamefont{Song}}\ and\ \bibinfo
  {author} {\bibfnamefont{G.}~\bibnamefont{Chen}},\ }%
  \bibfield{journal}{%
  \bibinfo {journal} {Appl. Phys. Lett.}\ }%
  \textbf{\bibinfo {volume} {84}},\ \bibinfo {pages} {687} (\bibinfo {year}
  {2004})\BibitemShut{NoStop}%
\bibitem{Hopkins_2009}%
  \BibitemOpen
  \bibfield{author}{%
  \bibinfo {author} {\bibfnamefont{P.~E.}\ \bibnamefont{Hopkins}}, \bibinfo
  {author} {\bibfnamefont{P.~T.}\ \bibnamefont{Rakich}}, \bibinfo {author}
  {\bibfnamefont{R.~H.}\ \bibnamefont{Olsson}}, \bibinfo {author}
  {\bibfnamefont{I.~F.}\ \bibnamefont{El-kady}},\ and\ \bibinfo {author}
  {\bibfnamefont{L.~M.}\ \bibnamefont{Phinney}},\ }%
  \bibfield{journal}{%
  \bibinfo {journal} {Appl. Phys. Lett.}\ }%
  \textbf{\bibinfo {volume} {95}},\ \bibinfo {pages} {161902} (\bibinfo {year}
  {2009})\BibitemShut{NoStop}%
\bibitem{Alvarez_2010}%
  \BibitemOpen
  \bibfield{author}{%
  \bibinfo {author} {\bibfnamefont{F.~X.}\ \bibnamefont{Alvarez}}, \bibinfo
  {author} {\bibfnamefont{D.}~\bibnamefont{Jou}},\ and\ \bibinfo {author}
  {\bibfnamefont{A.}~\bibnamefont{Sellitto}},\ }%
  \bibfield{journal}{%
  \bibinfo {journal} {Appl. Phys. Lett.}\ }%
  \textbf{\bibinfo {volume} {97}},\ \bibinfo {pages} {033103} (\bibinfo {year}
  {2010})\BibitemShut{NoStop}%
\bibitem{JustinNanotech}%
  \BibitemOpen
  \bibfield{author}{%
  \bibinfo {author} {\bibfnamefont{J.}~\bibnamefont{Haskins}}, \bibinfo
  {author} {\bibfnamefont{A.}~\bibnamefont{Kinaci}},\ and\ \bibinfo {author}
  {\bibfnamefont{T.}~\bibnamefont{Cagin}},\ }%
  \bibfield{journal}{%
  \bibinfo {journal} {Nanotechnology}\ }%
  \textbf{\bibinfo {volume} {22}},\ \bibinfo {pages} {155701} (\bibinfo {year}
  {2011})\BibitemShut{NoStop}%
\bibitem{JustinACS}%
  \BibitemOpen
  \bibfield{author}{%
  \bibinfo {author} {\bibfnamefont{J.}~\bibnamefont{Haskins}}, \bibinfo
  {author} {\bibfnamefont{A.}~\bibnamefont{Kinaci}}, \bibinfo {author}
  {\bibfnamefont{C.}~\bibnamefont{Sevik}}, \bibinfo {author}
  {\bibfnamefont{H.}~\bibnamefont{Sevin\ifmmode~\mbox{\c{c}}\else
  \c{c}\fi{}li}}, \bibinfo {author}
  {\bibfnamefont{G.}~\bibnamefont{Cuniberti}},\ and\ \bibinfo {author}
  {\bibfnamefont{T.}~\bibnamefont{\c{C}a\u{g}{\i}n}},\ }%
  \bibfield{journal}{%
  \bibinfo {journal} {ACS Nano}\ }%
  \textbf{\bibinfo {volume} {5}},\ \bibinfo {pages} {3779} (\bibinfo {year}
  {2011})\BibitemShut{NoStop}%
\bibitem{Cem-BN-PRB}%
  \BibitemOpen
  \bibfield{author}{%
  \bibinfo {author} {\bibfnamefont{C.}~\bibnamefont{Sevik}}, \bibinfo {author}
  {\bibfnamefont{A.}~\bibnamefont{Kinaci}}, \bibinfo {author}
  {\bibfnamefont{J.~B.}\ \bibnamefont{Haskins}},\ and\ \bibinfo {author}
  {\bibfnamefont{T.}~\bibnamefont{{\c C}a{\u g}{\i}n}},\ }%
  \bibfield{journal}{%
  \bibinfo {journal} {Phys. Rev. B}\ }%
  \textbf{\bibinfo {volume} {84}},\ \bibinfo {pages} {085409} (\bibinfo {year}
  {2011})\BibitemShut{NoStop}%
\end{thebibliography}

%

\end{document}